\documentclass[11pt]{iopart}

\expandafter\let\csname equation*\endcsname\relax
\expandafter\let\csname endequation*\endcsname\relax

\usepackage{graphicx}
\usepackage{bm}
\usepackage{color}
\usepackage{MnSymbol}
\usepackage{enumerate}
\usepackage{bbold}
\usepackage{lipsum}
\usepackage{array}
\usepackage{hyperref}
\usepackage{textgreek}
\usepackage{soul}
\usepackage{multirow}
\usepackage{accents}
\usepackage{trimclip}
\usepackage{tikz}
\usepackage{orcidlink}
\usepackage{comment}

\usepackage{pifont}
\newcommand{\cmark}{\ding{51}}%
\newcommand{\xmark}{\ding{55}}%

\DeclareMathAlphabet\mathbfcal{OMS}{cmsy}{b}{n}

\def\la{\langle}
\def\ra{\rangle}

\newcommand{\beq}{\begin{equation}}
\newcommand{\eeq}{\end{equation}}

\newcommand{\erf}[1]{Eq.~(\ref{#1})}

\newcommand{\dg}{^\dagger}

\newcommand{\op}[1]{\hat{ #1}}                

\newcommand{\an}[1]{\left\langle{#1}\right\rangle}

\renewcommand{\c}{_{\bo}}
\newcommand{\ie}{\textit{i.e.}}

\newcommand{\accentfuturearrowo}[1]{%
\begin{tikzpicture}[#1]%
\draw[line width = 0.2mm] (-0.6mm,0mm) -- (1.2mm,0mm);
\draw (2mm,0) -- (1.2mm,0.3mm) -- (1.2mm,-0.3mm) -- cycle;
\end{tikzpicture}%
}

\newcommand{\tempfuto}{\accentfuturearrowo{}}
\newcommand{\fut}[1]{\accentset{\tempfuto}{#1}}
\newcommand{\tempsfuto}{\scalebox{0.8}[0.8]{\accentfuturearrowo{}}}
\newcommand{\futs}[1]{\accentset{\tempsfuto}{#1}}

\newcommand{\accentfuturearrowc}[1]{%
\begin{tikzpicture}[#1]%
\fill (2mm,0) -- (1.2mm,0.3mm) -- (1.2mm,-0.3mm);
\draw[line width = 0.2mm] (-0.6mm,0mm) -- (1.2mm,0mm);
\draw (2mm,0) -- (1.2mm,0.3mm) -- (1.2mm,-0.3mm) -- cycle;
\end{tikzpicture}%
}

\newcommand{\temppast}{\scalebox{-1}[1]{\accentfuturearrowc{}}}
\newcommand{\past}[1]{\accentset{\temppast}{#1}}
\newcommand{\tempspast}{\scalebox{-0.8}[0.8]{\accentfuturearrowc{}}}
\newcommand{\pasts}[1]{\accentset{\tempspast}{#1}}

\newcommand{\accentbotharrowo}[1]{%
\begin{tikzpicture}[#1]%
\fill (-0.6mm,0) -- (0.2mm,0.3mm) -- (0.1mm,-0.3mm);
\draw[line width = 0.2mm] (-0.1mm,0mm) -- (1.2mm,0mm);
\draw (2mm,0) -- (1.2mm,0.3mm) -- (1.2mm,-0.3mm) -- cycle;
\draw (-0.6mm,0) -- (0.2mm,0.3mm) -- (0.2mm,-0.3mm) -- cycle;
\end{tikzpicture}%
}

\newcommand{\tempbotho}{\accentbotharrowo{}}
\newcommand{\both}[1]{\accentset{\tempbotho}{#1}}
\newcommand{\tempsbotho}{\scalebox{0.8}[0.8]{\accentbotharrowo{}}}
\newcommand{\boths}[1]{\accentset{\tempsbotho}{#1}}

\newcommand{\fil}{_{\text F}}
\newcommand{\sm}{_{\text S}}
\newcommand{\god}{_{\text T}}
\newcommand{\subx}{_{\text X}}
\newcommand{\suby}{_{\text Y}}
\newcommand{\subn}{_{\text N}}
\newcommand{\subp}{_{\Phi}}

\newcommand{\suba}{_{\text O}}
\newcommand{\subb}{_{\text U}}

\newcommand{\subss}{_\text{ss}}

\newcommand{\rhoOU}{\rho_{\pasts{\bo}, \pasts{\bu}}}
\newcommand{\rhoOV}{\rho_{\pasts{\bo}, \pasts{V}}}

\newcommand{\rhoO}{\rho_{\pasts{\bo}}}
\newcommand{\rhoOO}{\rho_{\boths{\bo}}}

\DeclareMathOperator*{\argmin}{arg\,min}

\newcommand{\dU}{^{\rm dU}}
\newcommand{\dV}{^{\rm dV}}
\newcommand{\dW}{^{\rm dW}}

\definecolor{nblue}{rgb}{0.06,0.3,0.73}
\definecolor{nblack}{rgb}{0,0,0}
\definecolor{nred}{rgb}{0.9,0.1,0.1}
\definecolor{nmagenta}{rgb}{0.7,0.0,0.3}
\definecolor{npurple}{rgb}{0.52,0,0.52}
\definecolor{ncyan}{cmyk}{1,0,0,0}
\definecolor{neditcolor}{rgb}{0.3,0.3,0.9}

\newcommand{\bo}{O}
\newcommand{\bu}{ U}
\newcommand{\dd}{{\rm d}}
\newcommand{\dt}{\dd t}

\newcommand{\EE}{{\rm E}}

\begin{document}

\title{Quantum state smoothing when Alice assumes the wrong type of monitoring by Bob} 

\author{Areeya Chantasri\orcidlink{0000-0001-9775-536X}$^{1,2}$, Kiarn T. Laverick\orcidlink{0000-0002-3688-1159}$^{2,3,4}$, Howard M. Wiseman\orcidlink{0000-0001-6815-854X}$^{2}$}
\address{$^1$Optical and Quantum Physics Laboratory, Department of Physics, Faculty of Science, Mahidol University, Bangkok, 10400, Thailand}%
\address{$^2$Centre for Quantum Computation and Communication Technology (Australian Research Council), Quantum and Advanced Technology Research Institute, Griffith University, Yuggera Country, Brisbane, QLD 4111, Australia}
\address{$^3$MajuLab, CNRS-UCA-SU-NUS-NTU International Joint Research Laboratory}
\address{$^4$Centre for Quantum Technologies, National University of Singapore, 117543 Singapore, Singapore}

\ead{areeya.chn@mahidol.ac.th, dr.kiarn.laverick@gmail.com, h.wiseman@griffith.edu.au}
\vspace{10pt}
\begin{indented}
\item[\today]
\end{indented}

\date{\today}

\begin{abstract}
An open quantum system leaks information into its environment. In some circumstances it is possible for an observer, say Alice, to recover that information, as a classical measurement record, in a variety of different ways, using different experimental setups. The optimal way for Alice to estimate the quantum state at time $t$ from the record before $t$ is known as quantum filtering. Recently, a version of quantum smoothing, in which Alice estimates the state at time $t$ using her record on both sides of $t$, has been developed. It requires Alice to make optimal inferences about the pre-$t$ record of a second observer, say Bob, who recovers whatever information Alice does not. But for Alice to make this inference, she needs to know Bob's setup. In this paper we consider what happens if Alice is mistaken in her assumption about Bob's setup. We show that the accuracy --- as measured by the Trace-Squared-Deviation, of Alice's estimate of the true state (\ie, the state conditioned on her and Bob's pre-$t$ records) --- depends strongly on her setup, Bob's actual setup, and the wrongly assumed setup. Using resonance fluorescence as a model system, we show numerically that in some cases the wrong smoothing is almost as accurate as the right smoothing, but in other cases much less accurate, even being less accurate than Alice's filtered estimate. Curiously, in some of the latter cases the {\em fidelity} of Alice's wrong estimate with the true state is actually higher than that of her right estimate. We explain this, and other features we observe numerically, by some simple analytical arguments.
\end{abstract}

\maketitle

\section{Introduction}

Many theoretical techniques applied to quantum systems have been borrowed from those successfully implemented in classical regimes. However, some have posed new puzzles, which could also lead to new discoveries. In this work, we are interested in problems of continuously monitored systems, where unknown physical states are to be estimated given partial information from observation. Two of the techniques in classical signal processing are filtering \cite{BookSpeyer,BookJazwinski,BroHwa12} and smoothing \cite{BookEinicke,BookWienerSmt,Sarkka13}, where noisy measurement signals can be `filtered' (causally) or `smoothed' (acausally) to estimate the underlining physical state dynamics of the system in time. The filtering technique can be readily extended to the quantum realm \cite{Bel87,BookCarmichael,Bel99,BookWiseman}. However, the extension of smoothing is less straight-forward, because it involves estimating a state using future information. This has led to various formalisms, including the two-state formalism~\cite{Watanabe1955,Watanabe1956,ABL1964} (generalized as the past quantum state formalism~\cite{Gammelmark2013,ZhaMol17}), weak values~\cite{AAV1988} and the weak-valued state~\cite{Gammelmark2013,Ohki22}, smoothed psuedo-probability distributions \cite{Tsangsmt2009,Tsangsmt2009-2} and the most likely path formalism~\cite{Chantasri2013,chantasri2015stochastic}. In this paper, we are concerned with another formalism, known as quantum state smoothing~\cite{Ivonne2015}. (See Ref.~\cite{ChaGue2021} for a unified presentation and application of many of these formalisms to an open quantum system.)

Quantum state smoothing applies to open quantum systems where information is missing. That is, the observer (Alice) has access to only some of the information leaking out of the system into its environment. Another observer (Bob), with access to the missing information as well as Alice's information, can estimate the system's properties better than Alice, allowing him to assign it a more pure state. Using quantum state smoothing, Alice can do better than using quantum state filtering, in the sense that she can assign a state that is closer to the unknown (to her, but known to Bob) state. This is a fair comparison because both quantum state smoothing and filtering can be defined as the task of optimally estimating the unknown state with the cost function being the trace-square-deviation~\cite{ChaGue2021} or, equivalently, the relative entropy~\cite{LavGueWis21}. The only difference is that quantum smoothing, like classical smoothing, uses Alice's future record as well as her past record.  

However, because of the differences between classical and quantum mechanics, quantum state smoothing has some differences from the classical case, even in situations where one might assume they are the same~\cite{LavWarWis23}.   Here, we are concerned with the fact that, contrary to classical state smoothing, the quantum version requires Alice to know {\em how} Bob is obtaining his information from the system's environment. That is, Bob has to choose a particular measurement setup to monitor the environment, and Alice has to know what that setup is. This difference raises intriguing questions about how the quantum state smoothing power varies with different setups of the unobserved (by Alice) channels~\cite{chantasri2019,LavCha2020a}. 
Here, by smoothing power, we mean how much improvement smoothing offers over filtering, in terms of the cost function. In fact, by choosing different setups, Bob changes his  `unraveling'~\cite{BookCarmichael} of the master equation, which can result in quantum smoothed states with quite different properties~\cite{LavWarWis23,chantasri2019,Laverick21}.

In this paper, we pose an even more dramatic question for the quantum state smoothing formalism: what would happen if Alice made a {\em wrong} assumption about what measurement setup Bob was using? How badly does this affect her ability to estimate the actual unknown state? Could smoothing even become worse than filtering? Does it depend on the characteristics of Bob's actual unraveling in relation to the wrongly assumed one? And does it depend on the characteristics of Alice's unraveling in relation to one or the other of these?

To address these questions, after detailing the quantum state smoothing formalism in Sec.~\ref{sec:QSS}, we introduce in Sec.~\ref{sec:model} the simple system we will focus upon. This is a driven two-level atom where three measurement setups are considered: photon detection, $x$-homodyne detection, and $y$-homodyne detection, as analyzed in Ref.~\cite{chantasri2019}. In Sec.~\ref{sec-conj}, we make some broad conjectures about smoothing power for all 18 combinations of Alice's setup, Bob's (actual) setup, and a (different) wrongly assumed setup for Bob. Numerical simulations in Sec.~\ref{sec:numerics} validate these conjectures, and also allow us to dig more deeply into the questions in the preceding paragraph. We find that with a wrongly assumed unobserved (Bob's) setup, the smoothing power (defined using the cost function) is sometimes roughly the same as that with the rightly assumed setup, but in other cases significantly worse than this, or even negative. This last case indicates that the quantum state smoothing is worse than the quantum state filtering (which is independent of Bob's measurement), but in this case,  paradoxically, the smoothed state typically has a higher fidelity with the unknown state than does the filtered state. We give some analytical explanations supplemented by numerical results for these observations in Sec.~6. Finally, we conclude in Sec.~\ref{sec-conclude}.

\section{Quantum State Smoothing} \label{sec:QSS}

Let us consider an open quantum system coupled to two independent baths (or environments) under the strongest Markov assumption~\cite{LiLi2018}. One of the baths is monitored by an observer, Alice, so that a measurement record $\boths{\bo}$ is observed by Alice. Here the double-headed arrow means a record over the full duration of the experiment, the interval $[t_{\rm i},t_{\rm f})$. The other bath is not monitored by Alice, but it is assumed to yield a measurement record $\boths{\bu}$ by virtue of a second observer labelled Bob. That is, $\boths{\bu}$ is unobserved by Alice, our primary observer. Bob may be an actual observer, or just a stand-in for the way information is robustly present in the environment due to decoherence~\cite{Schlosshauer2019,Kofman22,Zurek22,Strasberg23}. Therefore, Alice has access to only part of the classical information about the quantum system. 

It is useful to consider an observer who knows both $\boths{\bo}$ and $\boths{\bu}$. To save introducing a third observer with this role, we can assume that Bob knows Alice's record and the measurement setup used to obtain it, as well as his own. Then Bob has all the information and, if the initial state of the system is pure, his conditioned state at any time $t\geq t_{\rm i}$, denoted by $\rhoOU$, will also be pure. Here $\past\bo$ and $\past\bu$ are the records from time $t_{\rm i}$ up to time $t$. 
Thus $\rhoOU$ can be regarded as the \emph{true} state of the quantum system at time $t$, since no other observer can assign a more pure state. 

Alice's task, defined loosely for now, is to estimate the true state (Bob's state) at time $t$, 
using only her observed record $\boths{\bo}$. 
This could involve making inferences about what records Bob may have, and with what probabilities. 
 
The most straightforward technique for Alice, without any need to know what type of monitoring happens on Bob's side, is to compute her estimated state from the observed record using the \emph{quantum trajectory} approach~\cite{BookCarmichael,BookWiseman}, also called quantum state filtering~\cite{Bel87,Bel99}. The technique yields an estimated state $\rhoO$ at any time $t$, conditioned on (past) observed measurement results from the initial time $t_{\rm i}$ up to the time $t$. The pure state $\rhoOU$ introduced above is exactly this kind of filtered state, but there conditioned on both the observed ($\past\bo$) and unobserved ($\past\bu$) records. Let us define a little more formally the past observed record as $\past\bo \equiv  \{ o_s : s \in [t_{\rm i}, t) \}$, where $o_s$ denotes the observed meausurement outcome at time $s$, and similarly for $\past\bu$. 
Alice's filtered quantum state, $\rhoO(t)$, 
which has also been denoted $\rho\fil(t)$,
is mixed because it is conditioned only on partial information. Indeed, it is defined without reference to Bob's information $\pasts\bu$. Nevertheless it can be related to the state conditioned on the latter~\cite{Ivonne2015}: 
\begin{align}\label{eq-filter}
\rhoO(t) = \left\la \rhoOU(t) \right\ra_{\pasts{\bu} | \pasts{\bo}} \equiv \sum_{\pasts{\bu}} \wp(\pasts{\bu} | \pasts{\bo}) \rhoOU(t).
\end{align}
That is, it could be computed from considering all possible true states $\rhoOU$ and summing over the unobserved record using $\wp(\pasts{\bu} | \pasts{\bo})$. This last is a classical probability density function of past unobserved records conditioned on the past observed record, which can be obtained by classically filtering (processing in real time) the record $\past\bo$. 

However, as shown in Refs.~\cite{Ivonne2015,ChaGue2021}, Alice can do better than quantum state filtering by additionally using the remaining measurement results after the estimation time $\fut\bo \equiv  \{ o_s : s \in [t, t_{\rm f}) \}$ (future record) to help with the state estimation. This leads to the quantum state smoothing technique, where the filtered probability distribution $\wp(\pasts{\bu} | \pasts{\bo})$ is replaced by the smoothed probability distribution $\wp(\pasts{\bu} | \both{\bo})$. The key difference is that smoothing requires processing of the whole record $\both{O}$, which must necessarily be done after the fact. In other words, the estimated state becomes
\begin{align}\label{eq-smooth}
\rhoOO\dU(t) = \left\la \rhoOU(t) \right\ra_{\pasts{\bu} | \boths{\bo}} \equiv \sum_{\pasts{\bu}} \wp(\pasts{\bu} | \boths{\bo}) \rhoOU(t). 
\end{align}
This is called the smoothed quantum state, which has also been denoted $\rho\sm(t)$, 
conditioned on the whole (past and future) observed record. We note that, to calculate a smoothed quantum state, Alice is required to make an assumption about Bob's  measurement setup (unobserved by Alice), so we have introduced the superscript dU in Eq.~\eqref{eq-smooth} to emphasise that.
The relationship between the two classical probability distributions used in Eq.~\eqref{eq-filter} and Eq.~\eqref{eq-smooth} is
\begin{align}
\wp(\past{\bu}|\both{\bo}) & \propto \wp(\fut{\bo}|\past{\bu},\past{\bo})\wp(\past{\bu}|\past{\bo}), \label{Bayesrelate}
\end{align}
which is Bayes' theorem with the factor $\wp(\fut\bo)$  omitted, leading to proportionality rather than equality. The relation (\ref{Bayesrelate}) is useful for numerically evaluating the smoothed quantum state (\ref{eq-smooth}), which is a non-trivial procedure; for details, see Refs.~\cite{ChaGue2021,LavGueWis21,GueWis20}.

In the above, we said that smoothing gives a better estimate than filtering, but that is meaningless unless we define the cost function for evaluating estimates. In fact, both 
the filtered and smoothed states in Eq.~\eqref{eq-filter} and Eq.~\eqref{eq-smooth}, respectively, 
are optimal estimators for the expected Trace Square Deviation (TrSD) cost functions, but conditioned on different information. The TrSD, also known as the square of the Hilbert-Schmidt distance \cite{Ben17}, is defined as 
\beq 
{\cal S}[\rho,\sigma] = \Tr[(\rho - \sigma)^2].
\eeq
As shown in Ref.~\cite{ChaGue2021}, the filtered state can be defined as
\begin{align}\label{eq-filtercost}
\rhoO(t) = & \argmin_{\rho \in \mathfrak{S}} \left\la {\cal S}[\rho,\rhoOU(t)]\right \ra_{\pasts{\bu} | \pasts{\bo}},
\end{align}
the element of the set of unit-trace positive semi-definite operators $\mathfrak{S}$  with minimum expected TrSD to the true state, conditioned on the past observed record. Similarly, the smoothed state can be defined as
\begin{align}\label{eq-smoothcost}
\rhoOO\dU(t) = & \argmin_{\rho \in \mathfrak{S}} \left\la {\cal S}[\rho,\rhoOU(t)] \right \ra_{\pasts{\bu} | \boths{\bo}},
\end{align}
the estimator that minimizes the expected TrSD to the true state conditioned on the past-future observed record. Assuming (as we will throughout) that the true state $\rhoOU$ is pure, the minimal value of the expected TrSD (\ie, when using the optimal estimator), is given by the impurity $1 - {\cal P}[\rho_O(t)]$. Here $\bo$ indicates $\past\bo$ or $\both\bo$, depending on whether we are doing filtering or smoothing, and  the purity is defined as usual, via 
\beq
{\cal P}[\rho] = \Tr[\rho^2]\,.
\eeq

Of course, the TrSD is only one of many quantifiers of the difference between two quantum states. The justification for choosing it is that it is simple to use and reproduces the standard filtered state $\rhoO$ (and in particular its property of being independent of Bob's monitoring choice). In fact, the same filtered and smoothed states also arise from using relative entropy in place of TrSD~\cite{LavGueWis21}, but the TrSD is mathematically simpler. By contrast, using negative Fidelity --- with the negative sign allowing us to consider it a cost function, to be minimized --- instead of TrSD leads to optimal filtered and smoothed states that are different. Since we have assumed pure true states, the Fidelity, as defined originally by Jozsa~\cite{Jozsa1994}, can be simplified to 
\beq\label{eq-fid}
{\cal F}[\rho, \sigma] = {\rm Tr}\left[ \rho\, \sigma\right],
\eeq
which holds if either $\rho$ or $\sigma$ is pure. The optimal estimators by the negative Fidelity cost function are ``lustrated'' versions of the standard ones~\cite{ChaGue2021,LavGueWis21}. In this paper we stick to the TrSD as the cost function to define the estimates, but we will also calculate the Fidelity of these estimates, as we now discuss.

\subsection{Quantities for state comparisons  between valid and wrong smoothing}\label{sec-sp}

While the expected TrSD from the true state is the quantity that is minimized by the standard optimal smoothed (or filtered) quantum state (depending on the information available), there are other natural quantities that quantify the deviation from the true state, as discussed above.  
In particular, it is interesting to compute for our estimates not only the expected TrSD, but also the expected  Fidelity of the estimated state with the true state. The TrSD, as introduced  in the previous section, has values between $0$ (the two states are identical) and $2$ (the two states are pure and orthogonal). By contrast,  
the Fidelity ${\cal F}$ goes from $0$ to $1$, 
which are obtained when the two states are orthogonal and identical, respectively. 
Interestingly, it has also been proven~\cite{LavGueWis21,chantasri2019} that, for a given observed record, $O$, and an optimal estimate $\rho_O$ (in the TrSD sense defined above), 
\begin{equation}
     1-\an{{\cal S}[\rho_O\dU(t),\rhoOU(t)]}_{\past{U}|O} = \an{{\cal F}[\rho_O\dU(t),\rhoOU(t)]}_{\past{U}|O} = {\cal P}(\rho_O^{\rm dU}(t)) . \label{eq:equalitySFP}
\end{equation}
That is, the expected distance of the estimated state from the true state, by either TrSD or Fidelity measures, can be evaluated simply by computing 
the Purity of the (TrSD-optimal) state $\rho\c$, 
whether $\bo = \past{\bo}$ (filtering) or $\bo = \both{\bo}$ (smoothing). In previous work~\cite{ChaGue2021,chantasri2019}, it has been shown that smoothed states generally give higher purities, and hence a higher fidelity with the true states, which one would expect given that the smoothed state uses more information (the whole observed record) than the filtered state (past-only observed record).

However, in this paper, we are interested in the case of the sub-optimal estimates that arise when Alice makes an erroneous assumption about Bob's setup when computing her smoothed estimate. To understand the features of such ``wrong'' or ``erroneous'' smoothed quantum states, it is helpful to distinguish these cases, which we do by using the notation dV and $V$ for the valid setting and records, respectively,  and dW and $W$ for the wrong case, in place of dU and $U$. In the first case, Alice makes the valid assumption about Bob's measurement setup, and thus considers the correct possible records, leading to ``right smoothing'', with $\rhoOO\dV(t)$ given by Eq.~\eqref{eq-smooth} with dU $=$ dV and $U=V$. In the second case, Alice assumes the wrong setup, leading to ``wrong smoothing'', with $\rhoOO\dW(t)$ given by Eq.~\eqref{eq-smooth} with dU $=$ dW and $U=W$. This is ``wrong'' because the true state is still determined by $\past{V}$, not $\past{W}$, meaning that 
none of the equalities in Eq.~(\ref{eq:equalitySFP}) will hold: 
\begin{equation}\label{eq:equalitydW}
     {\cal P}(\rhoOO^{\rm dW}) \neq  1-\an{{\cal S}[\rhoOO\dW(t),\rhoOV(t)]}_{\pasts{V}|O} \neq \an{{\cal F}[\rhoOO\dW(t),\rhoOV(t)]}_{\pasts{V}|O}  \neq {\cal P}(\rhoOO^{\rm dW}) .
\end{equation}
For the rest of the paper, the notations dU (and U) will be used only when it can be replaced by either dV (and V) or dW (and W). 

Since the expected  distances between filtered or smoothed states and true states vary from realization to realization of the observed records, $\boths{O}$, just as the purity of the conditioned states vary. 
Thus, to make quantitative comparisons of the different scenarios around measurement setups and beliefs about them, we want to average  over all possible observed records as well. We denote this average, over both observed and  valid unobserved records, by E$[\cdot]$ (for total ensemble average). For the TrSD we have 
\beq
\EE[{\cal S}[\rho\c\dU(t),\rhoOV(t)] ] \equiv \sum_{{\bo},\past{V}} \wp({\bo},\past{V}) {\rm Tr}\left\{ [\rho\c\dU(t) - \rhoOV(t)]^2 \right\},
\eeq
as an average TrSD distance for the estimated state $\rho_{O}\dU$. 
Similar expressions hold for the ensemble average fidelity ${\cal F}$ and Purity ${\cal P}$, but we note that the latter does not require averaging over $\boths{V}$. 
In particular, we consider what would be measures of the power of smoothing relative to filtering, as follows.\\

\textit{1) Smoothing Power in terms of (negative) Trace Square Deviation}
\beq
{\cal R}_{{\cal S}}\dU(t) = - \EE[ {\cal S}[\rhoOO\dU(t),\rhoOV(t)]] + \EE[ {\cal S}[\rhoO(t),\rhoOV(t)]],
\label{eq-ATSDR}
\eeq

\textit{2) Smoothing Power in terms of Fidelity}
\beq
{\cal R}_{{\cal F}}\dU(t) = + \EE[ {\cal F}[\rhoOO\dU(t),\rhoOV(t)]] - \EE[ {\cal F}[\rhoO(t),\rhoOV(t)]],
\label{eq-AFR}
\eeq

\textit{3) Smoothing Power in terms of Purity}
\beq
{\cal R}_{{\cal P}}\dU(t) = + \EE[ {\cal P}[\rhoOO\dU(t)]] - \EE[ {\cal P}[\rhoO\dU(t)]].
\label{eq-APR}
\eeq\\
Note, the difference in sign for the TrSD, as also appearing in \erf{eq:equalitySFP}, is because it is a quantity we wish to minimize, while the Fidelity   and Purity are quantities we would expect (at least na\"ively) to be maximized for a good estimate. If Alice's assumption about Bob's measurement setup is valid, one would expect a positive smoothing power in all three cases. However, if Alice assumes the wrong setup, it is not obvious that this will be so. Before calculating what does happen, we introduce our model system (Sec.~\ref{sec:model}) and make some predictions (Sec.~\ref{sec-conj}).

\section{Model system: Coherently-driven qubit in vacuum bosonic baths} \label{sec:model}

Let us consider the example of a coherently-driven qubit, as in Refs.~\cite{Ivonne2015,ChaGue2021,LavGueWis21,chantasri2019}. The single qubit is driven by a qubit oscillation around the $x$-axis of the Bloch sphere with a frequency $\Omega$ and is coupled to two baths described by the Lindblad equation:
\beq
\dd \rho(t) = - i \dt [\hat{H} , \rho(t)] +  \dt\, {\cal D}[\op{c}_{\rm o}]\rho(t) + \dt\,{\cal D}[\op{c}_{\rm u}]\rho(t).
\label{eq-master}
\eeq
Here Alice only observes the first channel $\op{c}_{\rm o}$ and Bob measures the other bath $\op{c}_{\rm u}$ which is hidden from Alice (unobserved channel). The Hamiltonian is $\hat{H} = (\Omega/2) \,\op{\sigma}_x$ and the two Lindblad operators describing coupling to bosonic baths are assumed to be the (qubit fluorescence) lowering operators: $\op{c}_{\rm o,\rm u} = \sqrt{\gamma/2}\,\op{\sigma}_-$ with the same decay rate $\gamma/2$ such that both channels acquire the same amount of information.

Following Ref.~\cite{chantasri2019}, we consider three types of setup for the continuous measurement: (a) direct photon detection (leading to a jump-like unravelling), (b) $x$-quadrature homodyne detection (leading to a diffusive unravelling), and (c) $y$-quadrature homodyne detection (also diffusive). We use the symbol $\dd J$ for the measurement record obtained in a time step of duration $\dt$, with subscripts N, X, and Y to distinguish the three different types of monitoring. 

For the photon detection setup, the photon count in the infinitesimal interval is either $\dd J\subn = 1$ (click) or $\dd J\subn=0$ (no click). The respective operations~\cite{BookWiseman,BookKraus} (trace-decreasing maps) are ${\cal M}_{1} \rho = \op{c}\subn \, \rho \,  \op{c}\subn\dg {\rm d}t$ and ${\cal M}_0 \rho = M_0 \rho M_0\dg$, where $M_0 = {\hat 1} - i {\hat H} \dt - \tfrac{1}{2} \op{c}\subn\dg \op{c}\subn\dt$. Here $\op{c}\subn = \sqrt{\gamma/2} \op{\sigma}_-$, which is the same as $\op{c}_{\rm o, \rm u}$. 
The map ${\cal M}_{1}$ includes the unitary evolution described by the Hamiltonian $\op{H}$ but, for the sake of a simple presentation, omits the effect of the other channel. (Obviously when both channels are considered we do not include the Hamiltonian evolution twice;  this remark applies also to the diffusive cases below.)  Since $\dt$ is infinitesimal, the above dynamics can be described by the stochastic master equation for quantum jumps,
\beq\label{eq-smejump}
\dd \rho(t) = - i \, \dt [ \hat H , \rho(t)] +\dd J\subn(t) {\cal G}[\op{c}\subn] \rho(t) - \dt\, {\cal H}[\tfrac{1}{2} \op{c}\subn\dg \op{c}\subn] \rho(t),
\eeq
where we use~\cite{BookWiseman}
\begin{align}
{\cal G}[\op{c}] =&\,  \frac{ \op{c} \, \rho \, \op{c}\dg}{\Tr[\op{c} \, \rho \, \op{c}\dg]} - \rho, \\
{\cal H}[\op{c}] = &\, \op{c} \rho + \rho \op{c}\dg - \Tr[\op{c} \rho + \rho \op{c} ] \rho.
\end{align}

Turning now to the $x$- and $y$-homodyne detections, these are homodyne detections with local oscillator phases being $\Phi = 0$ and $\Phi = \pi/2$, respectively. In contrast to photon detection, homodyne measurement results during an infinitesimal interval can have any real value, denoted by $\dd J\subp$. Defining the Lindblad operators for these cases by $\op{c}\subp = \sqrt{\gamma/2} \op{\sigma}_- e^{-i \Phi}$, the associated measurement operations are ${\cal M}_{\dd J\subp} = M_{\dd J\subp} \rho M_{\dd J\subp}\dg$ where $M_{\dd J\subp} = {\hat 1} -i {\hat H} \dt - \tfrac{1}{2}  \op{c}\subp\dg \op{c}\subp \dt + \dd J\subp \op{c}\subp$. This yields a diffusive unravelling for the system state described by the stochastic master equation~\cite{BookWiseman},  
\beq\label{eq-smediff}
\dd \rho(t) = - i\, \dt [ \hat H , \rho(t)] + \dt \, {\cal D}[\op{c}\subp]\rho(t) + \dd W\subp(t) {\cal H}[\op{c}\subp]\rho(t).
\eeq
Here $\dd W\subp$ is a Wiener process related to the measurement record via $\dd W\subp(t) = \dd J\subp - \Tr[ (\op{c}\subp + \op{c}\subp\dg)\rho(t)] \dt $. For convenience, we denote $ \dd J\subx \equiv \dd J_{\Phi = 0} $ and $  \dd J\suby \equiv \dd J_{\Phi = \pi/2}$, for the results of $x$-homodyne and $y$-homodyne, respectively.

To ease our later discussion, we adopt the style of notation used in Ref.~\cite{chantasri2019}, generalized to the purpose of this paper: ``dO'' as shorthand for the ``Observed'' measurement setup or unravelling, and ``dU'' as shorthand for ``Unobserved'' measurement setup or unravelling. That is, we use the same capital letter for the monitoring setup as for the records themselves ($O$ and $U$) as in Sec.~\ref{sec:QSS}, but preceded by a ``d'' and all in Roman font. Also following Sec.~\ref{sec:QSS}, since we will allow for Alice to make either a valid or wrong assumption about Bob's measurement setup, the unobserved monitoring setup dU can be replaced by two options: ``dV'' for ``Valid'' (or right) unobserved setup and ``dW" for ``Wrong'' unobserved setup.

Each of dO, dV, and dW can be one of the three options: 
\begin{equation}
\begin{split}
{\rm dN}& \implies  \text{ jump records, $\dd J\subn$}, \\
{\rm dX} &\implies \text{ $x$-homodyne records, $\dd J\subx$}, \\
{\rm dY} & \implies \text{ $y$-homodyne records, $\dd J\suby$}.
\end{split}
\label{eq-def}
\end{equation}
For the right measurement records, there are in total nine combinations, denoted by dOdV = dNdN, dNdX, dNdY, dXdN, dXdX, dXdY, dYdN, dYdX, dYdY. However, each of the right combination can have two types of wrongly assumed setups dW (not to be confused with the Weiner increment $\dd W_\Phi$). Thus,  there are additional 18 combinations giving ``wrong smoothing''.

Before turning to the smoothed state under these 27 scenarios, we note that a wrong assumption by Alice about Bob's monitoring setup makes no difference to the optimal filtered state. That is because, as noted before, although the filtered state can be defined with reference to Bob's information $\pasts\bu$  as in \erf{eq-filter}, it can also be defined without any such reference. Thus, there are only 3 different types of filtered states, depending on whether Alice's setup dO equals dN, dX, or dY.

\section{Conjectures based on correlations of measurement records}\label{sec-conj}

From previous work~\cite{Ivonne2015,chantasri2019}, it has been shown that quantum state smoothing typically improves the fidelity between the estimated states to the hidden true states, relative to filtering. In particular,  Ref.~\cite{chantasri2019} investigated systematically the smoothing power in terms of Purity, ${\cal R}_{\cal P}$, in Eq.~(\ref{eq-APR}). This latter quantity is the average purity of the smoothed state minus that of the filtered states, and so in Ref.~\cite{chantasri2019} was called the average purity recovery. [Recall that in the case of a rightly assumed setup, ${\cal R}_{\cal P}$ is the same as the smoothing power in terms of Fidelity (${\cal R}_{\cal F}$) (\ref{eq-AFR}) and the smoothing power in terms of TrSD (${\cal R}_{\cal S}$) (\ref{eq-ATSDR}), because of \erf{eq:equalitySFP}.] Notably, 
it was found that there was a large variation in the smoothing power for different dOdV measurement setups (or dOdU, to use the notation of~\cite{chantasri2019}, where dOdW was not considered). More interestingly, it was shown  that this variation can be explained by the variation in correlation strength between the observed and unobserved measurement records. The greater the correlation strength, the higher the smoothing power. 

Now consider what will happen if Alice mistakenly assumes the wrong measurement setup for Bob when she does smoothing. We propose that the correlation strength should still have the ability to predict the smoothing power, because the correlation between two measurement records indicates how strong the stochastic measurement backaction on the system's state from one measurement affects the result of the other measurement. In the following, we propose four conjectures based on this intuition about correlation and measurement backaction, and how it would affect the power of (even wrong) smoothing. These conjectures were made prior to obtaining the numerical results in Sec.~\ref{sec:numerics}. 

\subsection{Correlations of measurement records}

Following Ref.~\cite{chantasri2019}, we begin by considering a normalized two-time correlation function between any two measurement records,  
\begin{align} \label{CFdef}
{\cal C}_2[{\rm dO, dU}]\equiv \frac{\EE[\dd J\suba(t+\tau)\dd J\subb(t)]\subss  - \EE[\dd J\suba(t+\tau)]\subss \EE[\dd J\subb(t)]\subss }{{\cal N}\suba {\cal N}\subb},
\end{align}
where $J\suba, J\subb \in \{ J\subn, J\subx, J\suby \}$, with $J_{\rm O}$ and $J_{\rm U}$ representing  the two simultaneous measurement signals, observed and unobserved by Alice respectively (recall that we have two output channels of equal strength in our model). 
The expected value $\EE[\cdot]\subss$ is computed over an ensemble of trajectories in the steady-state (ss) regime; that is, in any interval where the trajectory is statistically independent of the initial and final conditions. 
The normalized factors ${\cal N}$ in \erf{CFdef}, defined so that the correlations are all of the same magnitude and independent of $\dt$, are ${\cal N}_\bullet = \sqrt{\left\la (\dd J_\bullet)^2 \right\ra_{\rm ss} / \dt}$; see Ref.~\cite{chantasri2019} for more details. 
Unlike in that paper, here we consider only this two-point correlation function (as indicated by the subscript 2 of the ${\cal C}_2$ function in Eq.~\eqref{CFdef}), and  only the crudest classification of it, namely whether it is zero for all time differences $\tau$ or non-zero for some $\tau$. Note that the two-time correlators being zero does not imply that higher order (multi-time) correlations between the two measurement records dO and dU are zero~\cite{chantasri2019}. Nevertheless, the two-time correlation is the dominant contribution to the power of the smoothing, as was shown numerically in Ref.~\cite{chantasri2019}. Therefore, for the rest of the paper, we will use this notation without the `2' subscript: ${\cal C}[\rm dO, \rm dU] \sim $ \xmark\  to refer to ${\cal C}_2[\rm dO, \rm dU] = 0$ and ${\cal C}[\rm dO, \rm dU] \sim $ \cmark\  for ${\cal C}_2[\rm dO,\rm dU] \ne 0$.

\begin{table}[h]
\centering
\begin{tabular}{c|c|c|c|}
\cline{2-4}
& dX  & dY & dN \\ \hline
\multicolumn{1}{|c|}{dX} & \cmark\  & \xmark\ & \xmark\ \\ \hline
\multicolumn{1}{|c|}{dY} & \xmark\ & \cmark\ & \cmark\ \\ \hline
\multicolumn{1}{|c|}{dN} & \xmark\ & \cmark\ & \cmark\ \\ \hline
\end{tabular}
\caption{Summary of normalized two-time correlation functions between any two measurement records in a binary format: `\xmark' for zero correlation and `\cmark' for finite correlation. The measurement records are shown with the defined notation: dN, dX, and dY, referring to the photon count, the $x$-homodyne, the $y$-homodyne records, respectively. }
\label{tab-correlator}
\end{table}

The results for the two-time correlators among dN, dX, and dY are shown in Table~\ref{tab-correlator}. In words, the Table shows that: (1) the autocorrelations are always nonzero, as expected, (2) the cross correlation is nonzero only between dN and dY. This occurs because both dN and dY have measurement backaction in the $y$-$z$ plane of the Bloch sphere, which is in the same plane as the Rabi evolution, while dX is the only setup that gives the measurement backaction in the $x$-coordinate of the qubit~\cite{chantasri2019}. This   binary classification is sufficient because, for our system, the non-zero cross correlations are comparable in size to the autocorrelations~\cite{chantasri2019}.

\subsection{Four conjectures for valid-wrong smoothing}\label{sec-conjecture}

In Table~\ref{tab-conjecture}, we summarize our four conjectures regarding the smoothing power for a wrongly assumed setup, based on zero and non-zero correlators of the two types: ${\cal C}[{\rm dO,dV}]$ and ${\cal C}[{\rm dO,dW}].$
\begin{table}[h!]
\centering
{\renewcommand{\arraystretch}{1.3}
\renewcommand{\tabcolsep}{0.2cm}
\begin{tabular}{c|c|c|}
\cline{2-3}
& ${\cal C}[\rm dO,dV] \sim $ \xmark\   & ${\cal C}[\rm dO,dV]  \sim$ \cmark\   \\ \hline
\multicolumn{1}{|c|}{${\cal C}[\rm dO,dW] \sim $ \xmark\ } 

& {\bf C1}:  ${\cal R}\dW \sim {\cal R}\dV \sim $ small  &   {\bf C3}:  ${\cal R}\dW \ll {\cal R}\dV \sim $ large  \\ \hline
\multicolumn{1}{|c|}{${\cal C}[\rm dO,dW]  \sim$ \cmark\ } 
& {\bf C2}:  ${\cal R}\dW \sim $ ?, ${\cal R}\dV \sim $ small  & {\bf C4}:  ${\cal R}\dW \sim {\cal R}\dV \sim $ large   \\ \hline
\end{tabular}}
\caption{Summary of four conjectures (C1 - C4) based on the correlations between records  from Alice's setup (dO), records from Bob's actual setup (dV), and from a wrongly assumed setup for Bob (dW). The symbol ${\cal R}$ stands for all types of smoothing power defined in Section~\ref{sec-sp} except for the conjecture where a question mark appears (see text). In the main text ${\cal R}\dV$ and ${\cal R}\dW$ are referred to as right smoothing power and wrong smoothing power, respectively. Here ``large'' and ``small'' are relative terms.}
\label{tab-conjecture}
\end{table}

Let us start with the first two conjectures that are in the diagonal cells of Table~\ref{tab-conjecture}.\\

\textbf{Conjecture 1:} (Top-left of Table~\ref{tab-conjecture}) \textit{If the observed record is correlated neither with the unobserved record from the correct (valid) setup nor the record from the wrongly assumed setup, \ie, ${\cal C}[\rm dO, dV] \sim $ \xmark\  and ${\cal C}[\rm dO, dW] \sim $ \xmark, then the Wrong smoothing power should be roughly the same as the Valid smoothing power and both should be small.}

Cases for Conjecture 1: dOdVdW = dXdYdN, dXdNdY.\\

\textbf{Conjecture 4:} (Bottom-right of Table~\ref{tab-conjecture}) \textit{If the observed record is correlated both with the unobserved record from the correct (valid) setup and the record from the wrongly assumed setup, \ie, ${\cal C}[\rm dO, dV]  \sim$ \cmark\   and ${\cal C}[\rm dO, dW]  \sim$ \cmark\,  then the Wrong smoothing power should be roughly the same as the Valid smoothing power and both are not small.} 

Cases for Conjecture 4: dOdVdW = dYdYdN, dNdNdY, dYdNdY, dNdYdN\\

In the two conjectures above, the level of correlations between the observed and unobserved records are the same for both the correct (valid) setup (dV) and from the wrongly assumed setup (dW). The intuition behind these conjectures is as follows. In both case, the similar correlations between the observed record and the valid and wrong unobserved measurement setups implies that the different unobserved measurements have similar backaction on the system. As such, while the smoothed state will not be identical between the correctly and incorrectly assumed setups, we expect it to be similar. Thus, the performance in terms of smoothing power for the wrong smoothed state should be similar to that of the valid smoothed state, which is small for the uncorrelated case and large (in relative terms) for the correlated case. \\

\textbf{Conjecture 2:} (Bottom-left in Table~\ref{tab-conjecture}) \textit{If the observed record is uncorrelated with the correct (valid) unobserved record, \ie, ${\cal C}[\rm dO, dV] \sim $ \xmark, but it is correlated  with the record from the wrongly assumed setup, \ie, ${\cal C}[\rm dO, dW] \sim$ \cmark\, then the Valid smoothing power should be small, but the Wrong smoothing power may be strange.} 

Cases for Conjecture 2: dOdVdW = dXdNdX, dXdYdX, dNdXdN, dNdXdY, dYdXdY, dYdXdN.\\

In these cases, when there is near-zero correlation between  the records from dO and dV, it means that, to a first approximation~\cite{chantasri2019}, the unobserved measurement does not incur any measurement backaction onto the quantum system such that Alice's measurement can detect. That is, there should be little effect from Bob's measurement on Alice's observed record, and hence the right smoothing power will be small. In this situation, if Alice assumes an incorrect measurement setup dW that gives a record with a non-zero correlation with her observed record,  it is not obvious how the wrong smoothing power will behave, and the different measures of the power in Sec.~\ref{sec-sp} may be different.   
Specifically, since the smoothing with the valid setup assumption is optimal for the TrSD cost function (see Eq.~\eqref{eq-smoothcost}), the smoothing power, ${\cal R}_{\cal S}\dW$ of the wrong smoothing must be even worse than the right smoothing, ${\cal R}_{\cal S}\dV$. Thus ${\cal R}_{\cal S}\dW$ may even be negative; that is, the smoothing would gives a worse estimate than filtering. However, we cannot draw the same conclusion for the other types of smoothing power, ${\cal R}_{\cal F}$ and ${\cal R}_{\cal P}$, and for these the wrong smoothing may {\em appear} better than the valid smoothing. \\

\textbf{Conjecture 3:} (Top-right in Table~\ref{tab-conjecture}) \textit{If the observed record is correlated with the correct (valid) unobserved record, \ie, ${\cal C}[\rm dO, dV] \sim$ \cmark\, but uncorrelated with the record from the wrongly assumed setup, \ie, ${\cal C}[\rm dO, dW] \sim $ \xmark, then the Valid smoothing power will be large but the Wrong smoothing power will be small, if not zero.}

Cases for Conjecture 3: dOdVdW = dXdXdY, dXdXdN, dYdYdX, dNdNdX, dYdNdX, dNdYdX\\

The intuition for this conjecture comes from the fact that here, when Alice makes an incorrect assumption about Bob's measurement setup, she infers that his monitoring will  cause minimal backaction on the system. Thus Alice is effectively only taking into consideration her own measurement backaction when performing smoothing. As such, we can expect that this wrong smoothing will perform roughly as well as filtering.
That is, the smoothing power will be small at best, even though it is large for the valid setup. \\

\section{Numerical investigation} \label{sec:numerics}

In order to test our conjectures, we numerically generated qubit trajectories and analyzed the smoothing power using different measurement setups for the observed and unobserved channels. As discussed in Sec.~\ref{sec:model}, we are considering three types of measurement setups: dN, dX, and dY. 
This leads to 9 combinations of Alice's observed and Bob's (valid) unobserved measurement setups, dOdV, to consider. 
For each combination, the valid unobserved setup  (dV) can be mistakenly replaced by Alice, in her computations of the smoothed state, by two possible wrongly assumed setups (dW). Considering both valid and wrong smoothing, there are 27 cases in total.

For each of the 9 combinations of Alice’s
observed and Bob’s (valid) unobserved measurement setups, an ensemble of $3000$ true state trajectories ($\rho\god = \rho_{\pasts{O},\pasts{V}}$) is generated.  Each of these trajectories comes from a pair of observed ($\boths{\bo}$) and valid unobserved ($\boths{V}$) records, generated stochastically at the same time with the correct actual statistics. For each of the observed records, we numerically compute one filtered state trajectory, $\rhoO$, one valid smoothed state trajectory, $\rhoOO\dV$, and one wrongly-assumed smoothed state trajectory, $\rhoOO\dW$. For the smoothed state calculations, we follow Eq.~\eqref{eq-smooth}, where the ensemble averages were done over $10^4$ randomly-generated hypothetical unobserved records, given the appropriate Bob measurement setups  (valid in one case, wrong in the other). We note that for the smoothed states using the valid setups, the numerical data is replicated from Ref.~\cite{chantasri2019}. 
We use the same parameters defined in Ref.~\cite{chantasri2019}, i.e., the decay time $T_\gamma = 1/\gamma$ is set as the unit of time for all data plots, the qubit's unitary dynamics is an oscillation around the $x$-axis with the rate $\Omega = 5 \gamma$ and the measurement rate for both observed and unobserved channels is half of the total decay rate $\gamma/2$. The total time is $t_{\rm f} - t_{\rm i} = 8  \,T_\gamma$. More details of the calculation and numerical techniques for the simulation can be found in~\cite{Ivonne2015,chantasri2019}. 

\subsection{Examples of individual observed trajectories}

\begin{figure}[t]
\includegraphics[width=\textwidth]{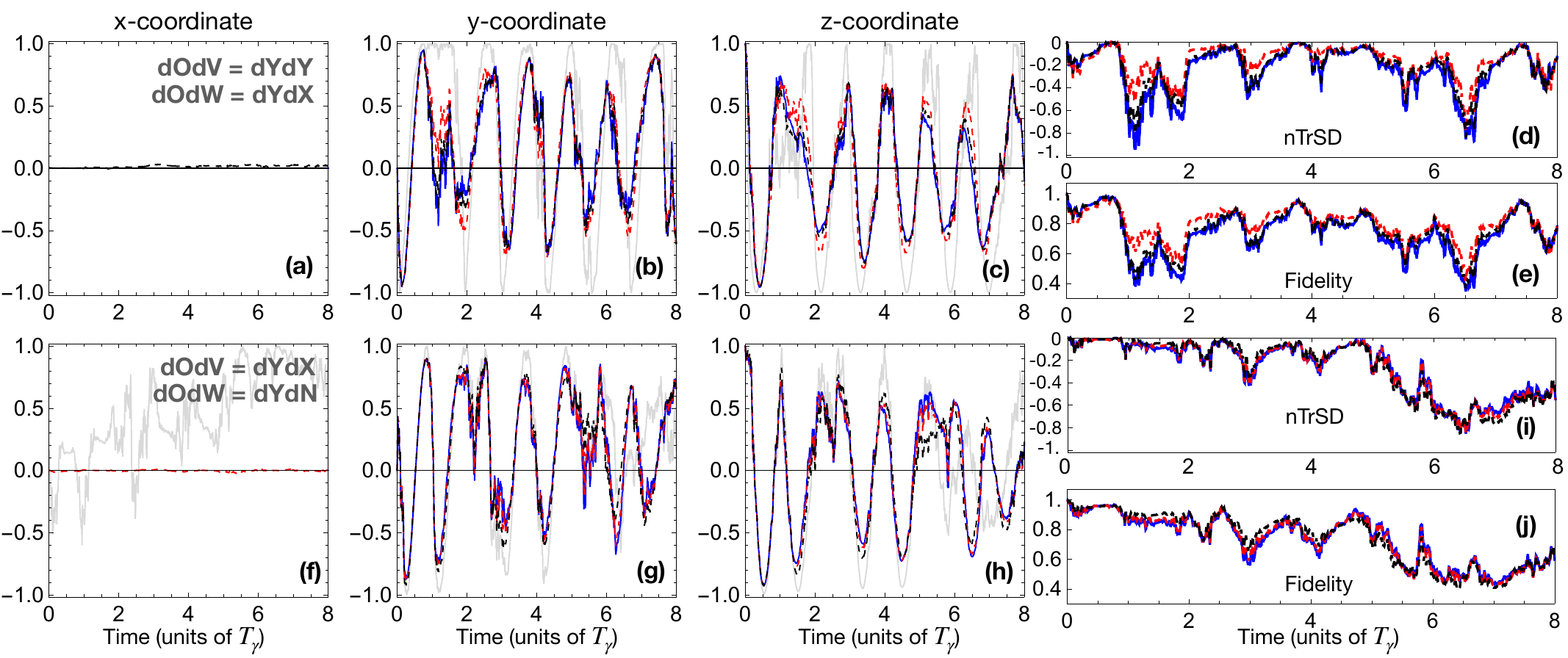}
\caption{Two examples of individual (true, filtered, and smoothed) trajectories and their corresponding nTrSD and Fidelity to true states. Panels in the first three columns, (a)-(c) and (f)-(h), show the Bloch sphere coordinates of: true state trajectories (light grey), filtered state trajectories (solid blue), valid smoothed state trajectories (dashed red), and wrongly-assumed smoothed state trajectories (dashed black). The last column shows the nTrSD and Fidelity between Alice's estimated states (filtering, valid smoothing, and wrong smoothing) and true states, using the same colour coding.}
\label{fig-example}
\end{figure}

We show in Figure~\ref{fig-example} examples of Alice's filtered (solid blue) and smoothed (dashed red) state trajectories and their corresponding negative Trace Square Deviation (nTrSD) and Fidelity to the true (grey) trajectories, for two combinations: dOdVdW = dYdYdX (top panels) and dOdVdW = dYdXdY  
(bottom panels). The two combinations are chosen such that obvious differences can be seen. 
The panels in the first three columns, (a)-(c) and (f)-(h), show the Bloch sphere coordinates of: true state trajectories, filtered state trajectories, valid smoothed state trajectories, and wrongly-assumed smoothed state trajectories. Because the observed records, dO = dY, are from the $y$-homodyne measurement in both examples, the filtered and smoothed states have vanishing $x$-components and finite oscillations in $y$- and $z$-coordinates. The non-zero $x$-component only occurs for the true state when the unobserved record, $\rm dV = dX$, is from the $x$-homodyne setup (shown in panel (f)).

Interesting results are shown in the last column of Figure~\ref{fig-example}, showing the negative TrSD and Fidelity (larger numbers mean better estimation of true states). The values fluctuate over time, but we can still see some trends. In the case of dOdV = dYdY (top panel), the valid smoothed state (dashed red) gives a better estimate of the true states than the one  with the wrongly assumed Bob setup dW = dX  (dashed black). In this case, the wrongly assumed smoothing still gives a better estimation quality than the filtered state (blue). However, in the case of dOdV = dYdX  with dW = dN (bottom panel),  the three curves are quite close, and often cross. This is an example of a case where we predicted the possibility of strange results.  A more systematic study is clearly needed in this case to discern the trends. 


\subsection{Trace-Square-Deviation (${\cal R}_{\cal S}$) and Fidelity (${\cal R}_{\cal F}$) smoothing  powers}\label{sec-powerresults} 

In this section, we analyse all 27 cases for the qubit examples numerically and determine whether the conjectures proposed in Section~\ref{sec-conjecture} are borne out for the 9 right smoothing cases (${\rm dU} = {\rm dV}$) and the 18 wrong smoothing cases (${\rm dU} = {\rm dW}$). 
From the smoothed state trajectories for different Alice's observed records, we compute their smoothing power, both in terms of TrSD, ${\cal R}_{\cal S}$, defined in Eq.~\eqref{eq-ATSDR}, and in terms of fidelity, ${\cal R}_{\cal F}$, defined in Eq.~\eqref{eq-AFR}. The smoothing power in terms of purity, ${\cal R}_{\cal P}$, defined in Eq.~\eqref{eq-APR}, is determined solely by the unobserved setup dU, regardless of whether it is the valid one or the wrong one. Thus ${\cal R}_{\cal P}$ is as already calculated in Ref.~\cite{chantasri2019}, and in this work is identical (for a sufficiently large ensemble) to ${\cal R}_{\cal S}^{\rm dU = dV}$ or ${\cal R}_{\cal F}^{\rm dU = dV}$. 

\begin{figure}[t]
\includegraphics[width=\textwidth]{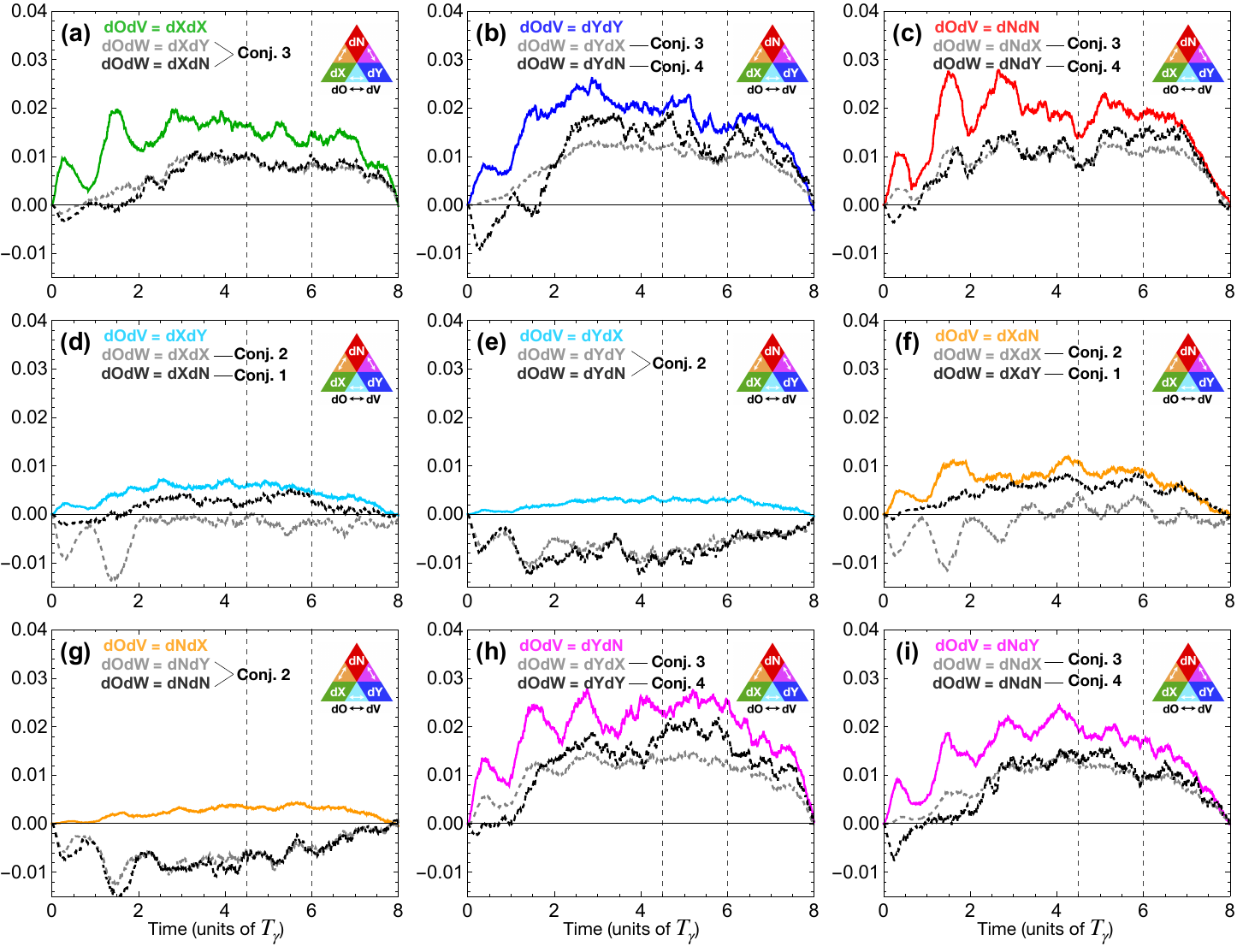}
\caption{Numerical results for the  negative Trace Square Deviation smoothing power (${\cal R}_{\cal S}$), defined in Eq.~\eqref{eq-ATSDR}, for all 27 cases, plotted as functions of time (units of $T_\gamma$), where the larger numbers mean better recoveries. The data of 27 cases are separated in 9 panels based on the 9 combinations of dOdV. The colour legends are used in reading the dO, dV, dW setups. The vertical dashed lines show the steady-state region where the transient behaviours from the initial and final conditions are minimal.  We also mark ``Conj.~1 - 4" to indicate the conjecture each curve correspond to.}
\label{fig-recovD}
\end{figure}

\begin{figure}[t]
\includegraphics[width=\textwidth]{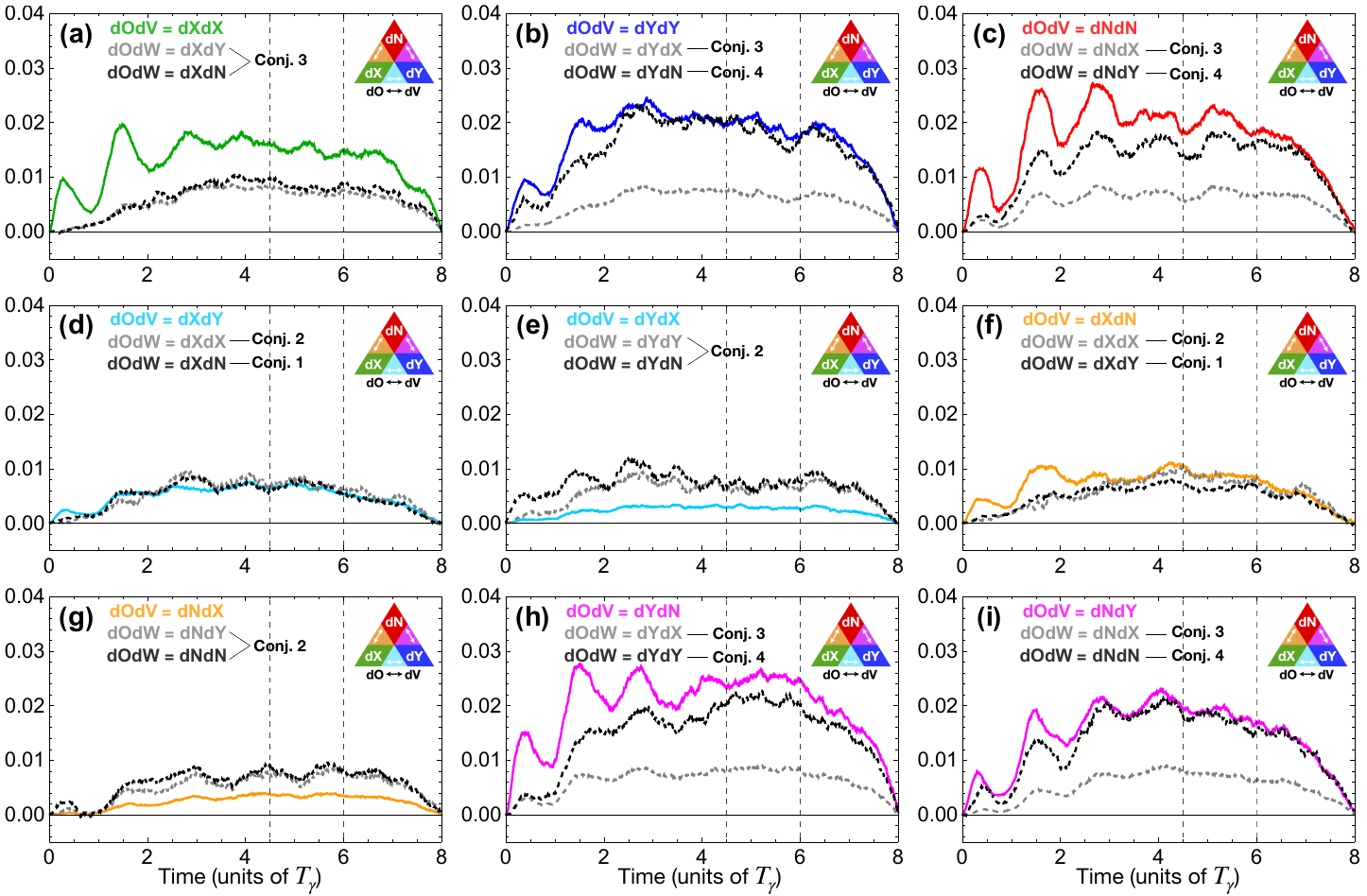}
\caption{Numerical results for the Fidelity smoothing power (${\cal R}_{\cal F}$), as defined in Eq.~\eqref{eq-AFR}. 
Other details are as in 
Figure~\ref{fig-recovD}. 
Importantly, the Purity smoothing power (${\cal R}_{\cal P}$), as it does not depend on whether the smoothing was valid or wrong, is exactly the Fidelity power and the TrSD power for the valid case, i.e., ${\cal R}_{\cal P}\dU = {\cal R}_{\cal F}^{\rm dU = dV} = {\cal R}_{\cal S}^{\rm dU = dV}$ for a given dO, and can be read from the colored (not gray nor black) curves.} 
\label{fig-recovF}
\end{figure}

All the numerical results for ${\cal R}_{\cal S}$ are shown in Figure~\ref{fig-recovD} and ${\cal R}_{\cal F}$ (also ${\cal R}_{\cal P}$) are shown in Figure~\ref{fig-recovF}. In each figure, there are 9 panels, covering all dOdU combinations, and each panel has 3 curves, corresponding to the valid smoothed state (${\rm dU} = {\rm dV}$, solid coloured curves) and the two more wrong smoothed states (${\rm dU} = {\rm dW} \neq {\rm dV}$, dashed black and grey curves). The triangle colour legend, which explains the colour given to dOdV, should be read according to the labels and the arrow, for example, dOdV = dXdX (green), dOdV = dYdN, dNdY (magenta), and dOdV = dNdX, dXdN (orange). We note that the recovery data is quite noisy because of the finite sample size as well as the transient behaviour at the start and end of the interval. 
The reader can focus on the results during the steady-state regime which is approximately $\tau_{\rm ss} = \{ t : t \in  [4.5\, T_\gamma, 6\, T_\gamma] \}$, marked by vertical dashed grey lines in the plots. We use this region to analyze all four conjectures, sumarized in Table~\ref{tab-conjecture}.

For Conjecture 1 in Table~\ref{tab-conjecture} (top left), both correlations, ${\cal C}[{\rm dO}, {\rm dV}]$ and ${\cal C}[{\rm dO}, {\rm dW}]$, are nearly zero, meaning that the observed record is not correlated, to lowest order, with any of the unobserved records (either righly or wrongly assumed). The prediction was that the wrong and valid smoothing powers should both be small. 
The numerical results corresponding to Conjecture 1 are the cases where dOdVdW = dXdYdN, dXdNdY, which are shown 
in panels (d) and (f) of Figures~\ref{fig-recovD} and \ref{fig-recovF}. We see that the wrong smoothing power (dashed black) are close to the right smoothing power (coloured solid), and both are quite small in comparsion to the smoothing powers in other panels, consistent with Conjecture 1.

For Conjecture 4 in Table~\ref{tab-conjecture} (bottom right), the opposite of Conjecture 1, both correlations, ${\cal C}[{\rm dO}, {\rm dV}]$ and ${\cal C}[{\rm dO}, {\rm dW}]$, are non-zero. 
The prediction was that the wrong and valid smoothing powers should both be not small.
We know from Ref.~\cite{chantasri2019} that ${\cal R}_{\cal F}\dV$ should be relatively large for the valid smoothing, and our prediction for the wrong smoothing is that ${\cal R}_{\cal F}\dW$ would be of similar size.  We can also extend the prediction to ${\cal R}_{\cal S}$ being relatively large in both the valid and wrong guessing cases.
There are four examples that correspond to Conjecture 4: dOdVdW = dYdYdN, dNdNdY, dYdNdY, dNdYdN, which are shown as dashed black curves in panels (b), (c), (h), and (i). In these panels of Figure~\ref{fig-recovD}, we indeed see that the ${\cal R}_{\cal S}$ of the wrongly assumed smoothed state (dashed black) is approximately as large as that of the valid smoothed state (solid coloured curves), albiet strictly lower. Also in Figure~\ref{fig-recovF}, the same panels, we see similar behaviour where ${\cal R}_{\cal F}$ of the wrong smoothing (dashed black) is quite similar to that of the valid smoothing (solid coloured curves). 

For Conjecture 2 in Table~\ref{tab-conjecture} (bottom left), the observed record is uncorrelated with the record from the valid Bob setup, ${\cal C}[{\rm dO}, {\rm dV}] \sim $ \xmark, but is correlated with that from the wrongly assumed one, ${\cal C}[{\rm dO}, {\rm dW}] \sim$ \cmark. This is where we may see big differences between the different types of smoothing power. For the valid smoothing, all types of smoothing power, ${\cal R}\dV$, are equal and should be small, as per Conjecture 1. For the wrong smoothing, the correlations suggest a large smoothing power.  However, we know that the ${\cal R}_{\cal S}\dW$ power must be less than ${\cal R}_{\cal S}\dV$ due to the optimality in Eq.~\ref{eq-smoothcost}, which means very small or even negative. But we cannot conclude this about ${\cal R}_{\cal F}$, allowing for a `strange' result.  This occurs for six combinations: dOdVdW = dXdYdX, dXdNdX, dYdXdY, dYdXdN, dNdXdY, and dNdXdN. The numerical results for the first two are shown as grey curves in panels (d) and (f) and the rest of the examples are shown as dashed black and grey curves in panels (e) and (g), in both figures. Interestingly, the negative values of ${\cal R}_{\cal S}\dW$ are obvious in these panels of Figure~\ref{fig-recovD}, especially in the panels (e) and (g). These negative values of ${\cal R}_{\cal S}\dW$ also somewhat coincide with the strange results of Fidelity power ${\cal R}_{\cal F}\dW$ in Figure~\ref{fig-recovF}(e) and (g), where the wrong smoothing power (dashed grey and black) can even be {\em larger} than the valid smoothing power (solid colored curves). These strange behaviours that the wrong smoothing could be worse than filtering in TrSD, i.e., ${\cal R}_{\cal S}\dW \lesssim 0$, but could be somehow better than valid smoothing in Fidelity, i.e., ${\cal R}_{\cal F}\dW > {\cal R}_{\cal F}\dV$, will be investigated further in Section~\ref{sec-furtherinvestigation}.

Finally, for Conjecture 3 in Table~\ref{tab-conjecture} (top right), the observed record is correlated with that from the valid Bob setup, ${\cal C}[{\rm dO}, {\rm dV}] \sim$ \cmark, but not correlated with that from the wrongly assumed one ${\cal C}[{\rm dO}, {\rm dW}] \sim $ \xmark\ . Here one clearly expects the valid smoothing power to always be greater than the wrong smoothing one, for both ${\cal R}_{\cal S}$ and ${\cal R}_{\cal F}$, with the wrongly assumed smoothing power being small. For the numerical results, there are six examples: dOdVdW = dXdXdY, dXdXdN, dYdYdX, dNdNdX, dYdNdX, dNdYdX. The first two are shown as dashed black and grey curves in panel (a), which are lower in smoothing power than the right smoothing one (solid green curve). The remaining four examples are shown as dashed grey curves in panels (b), (c), (h), and (i), which are also smaller than the right smoothing cases (solid coloured curves). We note that the above observations are more evident for ${\cal R}_{\cal F}$, in Figure~\ref{fig-recovF}, than for for ${\cal R}_{\cal S}$, in Figure~\ref{fig-recovD}. 
\\

\section{Refined analytical and numerical investigation}\label{sec-furtherinvestigation}

While the numerical results broadly support the four conjectures we made based on simple arguments, there is room both to make the arguments more rigorous and to explain details that emerged in the numerics which the conjectures did not specify. We begin by deriving some simple properties of quantum state smoothing power analytically.

To aid our analysis, let us write the smoothed states as
\begin{align}\label{eq-statedeviation}
    \rhoOO\dU =& \rhoO + \delta\rho_{\rm dU},
\end{align}
where $\delta\rho_{\rm dU}$ is traceless but need not be small.  
Then, from the fact that an ensemble of smoothed states with different future records (given a fixed past record) averages to the filtered state~\cite{LavCha2020a}, 
it follows  
that $\sum_{\fut{O}}\wp(\fut{O}|\past{O})\delta \rho_{\rm dU} = 0$. Using this (see \ref{App}), we can write the Fidelity and Purity smoothing powers in terms of $\delta\rho_{\rm dV}$ and $\delta\rho_{\rm dU}$ as
\begin{subequations}\label{eq-powerdelrho}
    \begin{align}
    {\cal R}_{\cal F}\dU =&\, {\rm E}[\Tr[\delta\rho_{\rm dU}\delta\rho_{\rm dV}]],\\
    {\cal R}_{\cal P}\dU =&\, {\rm E}[\Tr[(\delta\rho_{\rm dU})^2]] ,
\end{align}
\end{subequations}
which  apply to both dU = dV and dU = dW. For the TrSD power, we use a relationship (see \ref{App}) among the three smoothing powers to get
\begin{subequations}\label{eq-powerrelate}
\begin{align}
{\cal R}_{\cal S}\dU =& \, 2\,{\cal R}_{\cal F}\dU - {\cal R}_{\cal P}\dU,\\
=& \, 2\, {\rm E}[\Tr[\delta\rho_{\rm dU}\delta\rho_{\rm dV}]] - {\rm E}[\Tr[(\delta\rho_{\rm dU})^2]].
\end{align}
\end{subequations} 
Given these general definitions of the smoothing power in Eqs.~\eqref{eq-powerdelrho} and \eqref{eq-powerrelate}, we can show that, for the valid smoothing, all types of smoothing power are equal: 
\beq
{\cal R}_{\cal S}\dV = {\cal R}_{\cal F}\dV = {\cal R}_{\cal P}\dV  = {\rm E}[\Tr[(\delta\rho_{\rm dV})^2]],
\eeq 
as has been confirmed both analytically and numerically in Refs.~\cite{Ivonne2015,LavGueWis21,GueWis20}.

We now use the above properties and formulas for the smoothing powers to refine our explanation and predictions for the four conjectures. Firstly, for the Purity power, because the purity does not depend on the unknown true states, the power ${\cal R}_{\cal P}$ should monotonically depend on the correlation strength between Alice's observed and assumed unobserved records, ${\cal C}[\rm dO, \rm dU]$, irrespective of whether ${\rm dU}={\rm dV}$ or dW. That is, we conclude that
\begin{subequations}\label{eq-interpret1}
\begin{align}
        {\cal C}[{\rm dO}, {\rm dU}] \sim {\text \xmark}\ \,\, &\,\, \implies \,\,\, \,\, {\cal R}_{\cal P}\dU = {\rm E}[\Tr[(\delta\rho_{\rm dU})^2]] \textrm{  is small,} \\ 
        {\cal C}[{\rm dO},{\rm dU}] \sim {\text \cmark}\ \,\, & \,\, \implies \,\, \,\, {\cal R}_{\cal P}\dU = {\rm E}[\Tr[(\delta\rho_{\rm dU})^2]]  \textrm{  is larger.}
\end{align}
\end{subequations}
For the valid smoothing, since all types of smoothing power are equal, Eqs.~\eqref{eq-interpret1} are true when replacing ${\cal R}_{\cal P}\dU$ with ${\cal R}\dV$. 
These properties consistently explain the numerical results for Conjecture~1, 3 and 4.

We now focus solely on understanding the strange results of Conjecture~2, i.e., when the valid two-time correlation vanishes, ${\cal C}[{\rm dO},{\rm dV}] \sim $ \xmark, and the wrongly assumed correlation does not, ${\cal C}[{\rm dO},{\rm dW}] \sim $ \cmark.
The numerical results were clearly shown in Section~\ref{sec-powerresults} that the TrSD power can even be negative, i.e., ${\cal R}_{\cal S}\dW \lesssim 0$, and that the Fidelity power for the wrong guessing can be better than the valid one, i.e., ${\cal R}_{\cal F}\dW > {\cal R}_{\cal F}\dV$. These strange behaviours are particularly pronounced for some combinations of the measurement setups for Conjecture 2, i.e., dOdVdW = dYdXdY, dYdXdN, dNdXdY, dNdXdN, as shown in panels (e) and (g) of Figures~\ref{fig-recovD} and Figures~\ref{fig-recovF}. To investigate this, let us introduce another quantity that can be thought of as a correlation coefficient between the state deviations $\delta\rho_{\rm dV}$ and $\delta\rho_{\rm dW}$,
\begin{align}\label{eq-powercorr}
\alpha \equiv&\, 
\frac{{\cal R}_{\cal F}\dW}{\sqrt{{\cal R}_{\cal P}\dW {\cal R}_{\cal P}\dV}} 
=  \, \frac{ {\rm E}[\Tr[\delta\rho_{\rm dW}\delta\rho_{\rm dV}]] }{  \sqrt{{\rm E}[\Tr[(\delta\rho_{\rm dW})^2]]{\rm E}[\Tr[(\delta\rho_{\rm dV})^2]]}}.
\end{align}
Given that all smoothed states, regardless of the valid or wrong unobserved unravelling, are calculated with the same future observed record, $\fut{O}$, one would expect that the state deviation $\delta\rho_{\rm dU}$ for different types of unravelling dU should be highly correlated. That is, we expect that $\alpha$ should be 
of order 1, or even close to 1. 
This intuition is borne out by numerical simulations for all 27 cases of different dV and dW, which show that $\alpha$ is, in all cases, larger than $1/\sqrt{2}$ for the great majority of the time, and often much closer to $1$  (see Figure~\ref{fig-powerrelate} in \ref{sec-alpha}).

We now return to the two quantities we are particularly interested in for Conjecture 2 for wrongly assumed smoothing. These are the TrSD power and the Fidelity power compared to the Fidelity power for the valid smoothing. These can be written in terms of the above quantities, in Eqs.~\eqref{eq-powerdelrho}, \eqref{eq-powerrelate} and \eqref{eq-powercorr}, as follows: 
\begin{align}\label{eq-conj2exp1}
    {\cal R}_{\cal S}\dW & = 2 \, \alpha \sqrt{{\rm E}[\Tr[(\delta\rho_{\rm dW})^2]]{\rm E}[\Tr[(\delta\rho_{\rm dV})^2]]} - {\rm E}[\Tr[(\delta\rho_{\rm dW})^2], \\ 
    {\cal R}_{\cal F}\dW-{\cal R}_{\cal F}\dV &= \alpha \sqrt{{\rm E}[\Tr[(\delta\rho_{\rm dW})^2]]{\rm E}[\Tr[(\delta\rho_{\rm dV})^2]]} 
- {\rm E}[\Tr[(\delta\rho_{\rm dV})^2]. \label{eq-conj2exp2}
\end{align}
Thus it follows that both counterintuitive results --- the TrSD power of the wrong smoothing being negative (Eq.~\ref{eq-conj2exp1} being negative) 
and the Fidelity power of the wrong smoothing being larger than that of the valid smoothing (Eq.~\ref{eq-conj2exp2} being positive) --- occur at the same time if and only if the following is true: 
\begin{align}\label{eq-strangecondition}
     \frac{{\rm E}[\Tr[(\delta\rho_{\rm dV})^2]}{{\rm E}[\Tr[(\delta\rho_{\rm dW})^2]} < \min \left\{\alpha^2, \frac{1}{4 \alpha^2} \right\}.
\end{align}
Since $\min\{\alpha^2, 1/4\alpha^2\} \leq 1/2$, a {\em necessary} condition for Eq.~\eqref{eq-strangecondition} to be satisfied is for ${\cal R}_{\cal P}\dV$ to be at most half the size of ${\cal R}_{\cal P}\dW$, which only occurs for Conjecture~2 conditions, where we have ${\cal C}[{\rm dO},{\rm dV}] \sim $ \xmark\  and ${\cal C}[{\rm dO},{\rm dW}] \sim $ \cmark. 

But we can say more. 
Since the correlation between the state deviation is, in actuality, quite strong, i.e., $1/2 \le \alpha^2 \le 1$, then we have  $1/4 \leq \min\{\alpha^2, 1/4\alpha^2\} \leq 1/2$. That is, for our system, a {\em sufficient} condition for Eq.~\eqref{eq-strangecondition} to be satisfied, and so for 
both the counter-intuitive results to occur, is that 
the valid Purity smoothing power to be at most one quarter of the wrong Purity smoothing power. This is exactly where we see the clear manifestation of the above counterintuitive phenomena in panels (e) and (g) of Figures~\ref{fig-recovD} and \ref{fig-recovF}.
To verify that valid Purity smoothing power is at
most one quarter of the wrong Purity smoothing power, the reader can refer to the coloured curves for ${\cal R}_{\cal P}\dU$ for dU = dV or dW (which equals ${\cal R}_{\cal S}\dV$ in Figure~\ref{fig-recovD} or ${\cal R}_{\cal F}\dV$ in Figure~\ref{fig-recovF}). For example, comparing the cyan curve in panel (e), for ${\cal R}_{\cal P}^{\rm dV = dX}$, and the blue curve in panel (b), for ${\cal R}_{\cal P}^{\rm dW = dY}$, the condition is satisfied to explain the counter-intuitive result of dOdW = dYdY (the grey curve in panel (e)). Another example is to compare the orange curve in panel (g), for ${\cal R}_{\cal P}^{\rm dV = dX}$, and the magenta curve in panel (i), for ${\cal R}_{\cal P}^{\rm dW = dY}$, the condition is satisfied for the strange result of dOdW = dNdY (the grey curve in panel (g)).

\section{Conclusion}\label{sec-conclude}

In this paper, we have considered ``perplexing" quantum state smoothing scenarios, where an observer, Alice, who has access to only some of the information of a quantum system of interest, attempts to estimate its true state by conditioning on both her past and future records, 
but makes an erroneous assumption about how the missing quantum information was recovered by a hidden observer Bob. That is, the true states, conditioned on Alice's records and the hidden records arising from Bob's ``valid'' measurement scheme, are different from the ones Alice guesses because she assigns a ``wrong'' measurement scheme to Bob. 

We have analysed the power of Alice's smoothing, considering three different measures: the negative Trace Square Deviation (nTrSD) from the true state, Fidelity with the true state, and Purity. The first and second of these are actually averages over the valid unobserved (Bob's) record. By the power of quantum state smoothing we mean how much it increases these measures compared to filtering, the conventional state estimate using only past information. We evaluated these three measures for the same state estimates (one smoothed and one filtered), which are the optimal ones according to the nTrSD measure.

As a model system, we considered a two-level atom undergoing resonance fluoresence, where Alice and Bob could choose from three different measurement setups ---
photon detection, $x$-homodyne, and $y$-homodyne --- and thus in each case Alice could be wrong about Bob's measurement in two different ways. Guided by the strength of the correlations between Alice's observed records and Bob's records (unobserved by Alice), we constructed  conjectures for the various smoothing powers in four different regimes.
All of these conjectures were then validated through numerical simulations.  
The most interesting of the four conjectures related to the situation where Alice's record is highly correlated with the type of record she wrongly assumes Bob has, but poorly correlated with the actual type of record Bob has. 
Here we predicted the potential for ``strange'' results, in a manner we now explain. 

One might have guessed that Alice making a wrong assumption about the unobserved measurement setup would result in worse estimate of the true state --- by any measure --- than if she assumed the valid setup. Moreover, it might have been intuitive that, at the worst, the wrongly assumed smoothing should perform as well as the quantum state filtering, which uses the past-only measurement records. However, we predicted, and verified, that these intuitive guesses are not always true. The wrongly assumed smoothing in certain cases can result in higher average fidelity to true states than that of the correct (valid) smoothing. In those same cases, the wrongly assumed smoothing gives even lower average trace-square distance to true states than that of the conventional filtering. By analytical arguments supplemented by numerical simulations, we also found  necessary conditions and sufficient conditions for these strange behaviours to occur. We emphasise that, however strange, these behaviours are consistent with the optimality of all valid (not wrong) estimates with respect to the nTrSD measure.

In this paper we based our conjectures about the performance of the wrongly assumed quantum state smoothing only upon 
the two-time correlators of measurement records. One could, however, extend the analysis to consider higher-order correlators between the measurement records, in particular a three-time correlator as considered in Ref.~\cite{chantasri2019}. This would likely enable better quantitative predictions of the smoothing power for different combinations of settings (observed, valid, and wrongly assumed), and especially of when the sufficient conditions for ``strange'' behaviour would be met. 
Our consideration of wrongly assumed measurement settings also suggests a more exotic avenue for future work, namely that of counterfactual measurement settings. 
For example, one could consider the scenario where Alice performs $y$-homodyne detection, obtaining a record $\both{O}$, while she also knows what measurement Bob performs.  Given all of this information, Alice could consider the smoothed state she {\em would have} calculated if she had instead performed a different measurement, say $x$-homodyne.  Last but not least, it could also be interesting to investigate whether anything similar to the effects we have found occur 
in state smoothing for classical systems with wrong assumptions about the process noise model.

\ack
This work was supported by the Australian Research Council via the Centre of
Excellence grant CE170100012. A.C. also acknowledges the support of the Program
Management Unit for Human Resources and Institutional Development, Research and
Innovation (Thailand) grant B39G680007.

\appendix
\section{Derivation for Fidelity and TrSD smoothing powers}\label{App}
We begin by explicitly writing the Fidelity power for an unobserved measurement setup dU (which can be dV or dW) in the form of weighted sums, 
\begin{align}
{\cal R}_{\cal F}\dU &= {\rm E}[{\cal F}[\rhoOO\dU,\rhoOV] - {\cal F}[\rhoO,\rhoOV]] \\
& = {\rm E}[\Tr[\rhoOO\dU\rhoOV] - {\rm E}[\Tr[\rhoO\rhoOV]] \\
&= \sum_{\boths{O},\pasts{V}} \wp(\both{O},\past{V})\Tr[\rhoOO\dU\rhoOV] - \sum_{\boths{O},\pasts{V}} \wp(\both{O},\past{V})\Tr[\rhoO\rhoOV]\\
&= \sum_{\boths{O}} \wp(\both{O}) \left[\sum_{\pasts{V}} \wp(\past{V}|\both{O})\Tr[\rhoOO\dU\rhoOV]\right] - {\rm E}[{\cal P}(\rhoO)]\\
&= \sum_{\boths{O}} \wp(\both{O}) \Tr[\rhoOO\dU\rhoOO\dV] - {\rm E}[{\cal P}(\rhoO)] \label{eq-appfidel},
\end{align}
using the fidelity definition in Eq.~\eqref{eq-fid}. Substituting the smoothed state Eq.~\eqref{eq-statedeviation} in Eq.~\eqref{eq-appfidel} above, we get
\begin{align}
{\cal R}_{\cal F}\dU =& \sum_{\boths{O}} \wp(\both{O}) \Tr[(\rhoO+\delta\rho_{\rm dU})(\rhoO+\delta\rho_{\rm dV})] - {\rm E}[{\cal P}(\rhoO)]\\
=&\sum_{\boths{O}} \wp(\both{O}) \Tr[2\rhoO\delta\rho_{\rm dU} +\delta\rho_{\rm dU}\delta\rho_{\rm dV}] \\
=& \, \label{eq-RFdelta} {\rm E}[\Tr[\delta\rho_{\rm dU}\delta\rho_{\rm dV}]],
\end{align}
where we have applied the property that an ensemble of smoothed states averages back to a filtered state~\cite{LavCha2020a}, i.e.,
\beq
\sum_{\futs{O}}\wp(\fut{O}|\past{O}) \rhoOO\dU = \rhoO\,.
\eeq
which is equivalent to $\sum_{\fut{O}}\wp(\fut{O}|\past{O})\delta \rho_{\rm dU} = 0$ mentioned in the main text. Note also that the similar derivation is applied for the Purity power: 
\begin{align}
{\cal R}_{\cal P}\dU =&\, {\rm E}[{\cal P}[\rhoOO\dU] -{\rm E}[{\cal P}[\rhoO]] \\
=& \sum_{\boths{O}} \wp(\boths{O}) \Tr[(\rhoO + \delta \rho_{\rm dU})(\rhoO + \delta \rho_{\rm dU})] - {\rm E}[{\cal P}[\rhoO]] \\
= &\,  \sum_{\boths{O}} \wp(\boths{O}) \Tr[2 \delta \rho_{\rm dU}\,\rhoO + (\delta \rho_{\rm dU})^2]\\
=& \, \label{eq-RPdelta} \sum_{\boths{O}} \wp(\boths{O}) \Tr[(\delta \rho_{\rm dU})^2] \equiv {\rm E}[\Tr[(\delta\rho_{\rm dU})^2]].
\end{align}

For the TrSD power, let us first derive the relationship among the three smoothing power. The TrSD for a conditional state $\rho_{O}$ can be written as
\begin{align}
{\cal S}[\rho_{O},\rhoOV]&=\Tr\left[(\rho_{O} - \rhoOV)^2\right] \\
&= \Tr[\rho_{O}^2] - 2\Tr[\rho_{O}\rhoOV] + \Tr[\rhoOV^2] \\
&= {\cal P}[\rho_{O}] -2 {\cal F}[\rho_{O},\rhoOV] + 1\,,
\end{align}
where, in the final equality, we have assumed that the state $\rhoOV$ is pure, leading to the equality ${\cal F}[\rho_{O},\rhoOV] = \Tr[\rho_O \rhoOV]$. Importantly, this relationship holds irrespective of the choice of $\rho_{O}$, being $\rhoO$, $\rhoOO\dV$ or $\rhoOO\dW$. By taking an ensemble average of ${\cal S}[\rho_{O},\rhoOV]$ and substituting ${\cal R}_{\cal F}$ Eq.~\eqref{eq-RFdelta} and ${\cal R}_{\cal P}$ Eq.~\eqref{eq-RPdelta}, we get
\begin{align}
{\cal R}_{\cal S}\dU =& \, 2\,{\cal R}_{\cal F}\dU - {\cal R}_{\cal P}\dU\\
= & \, 2\,{\rm E}[\Tr[\delta\rho_{\rm dU}\delta\rho_{\rm dV}]] - {\rm E}[\Tr[\left(\delta\rho_{\rm dU}\right)^2]],
\end{align}
as shown in Eq.~\eqref{eq-powerrelate} in the main text.

\section{Numerical results of correlation coefficient between state deviations}\label{sec-alpha}

\begin{figure}[h]
\includegraphics[width=\textwidth]{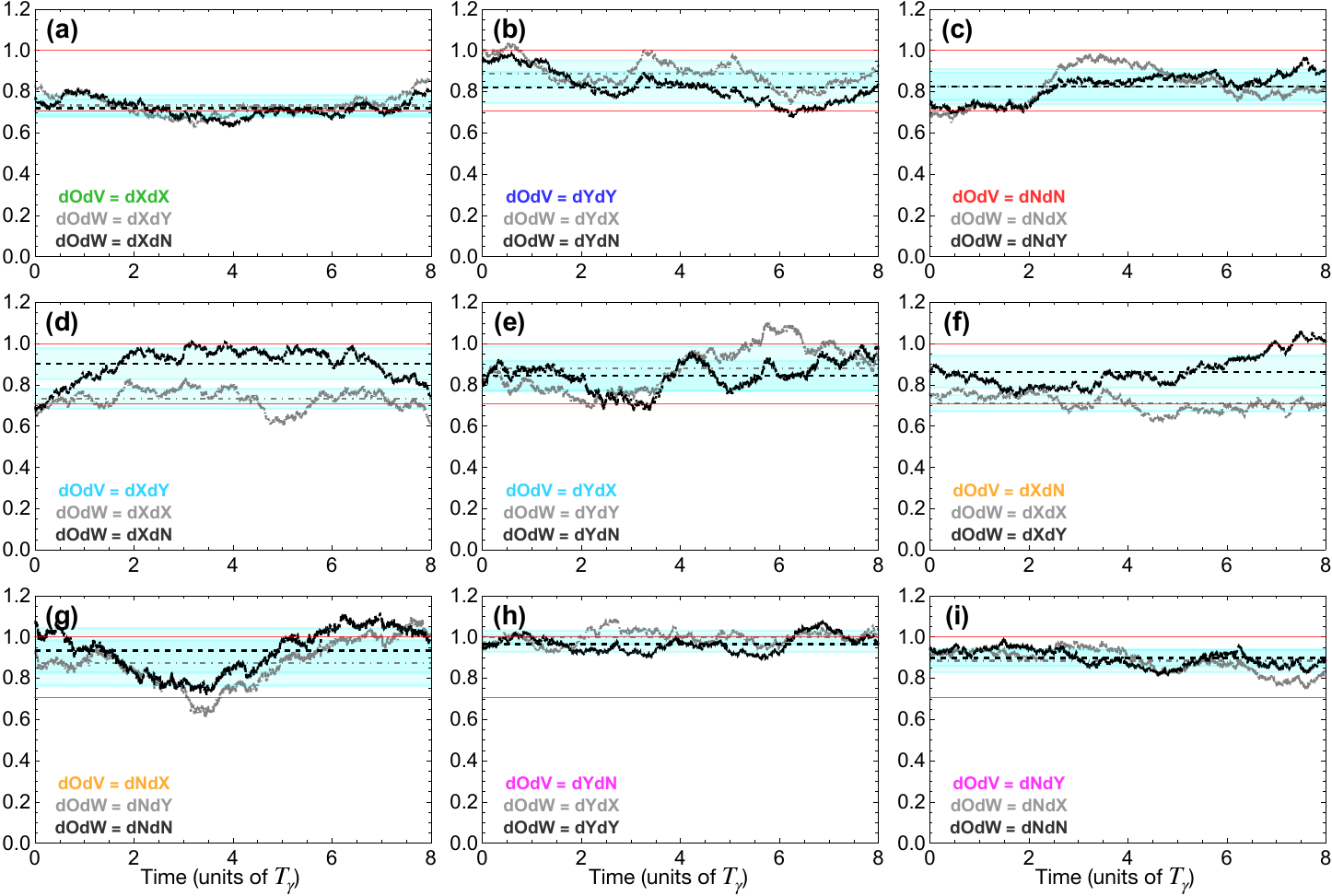}
\caption{Numerical results for the correlation coefficient $\alpha$ in Eq.~\eqref{eq-powercorr} or Eq.~\eqref{eq-appcorr}.
showing that it is always large for all combinations of observed and unobserved measurement setups. Each panel includes two curves (grey and black) corresponding to two different types of dW for each dOdV. The dOdVdW labels are replicated from Figures~\ref{fig-recovD} and \ref{fig-example} for convenience. The dotdashed grey and dashed black lines are the time-average values of the fluctuating grey and black curves, respectively. The cyan shades show the area of one standard deviation from the averages. The red lines show $\alpha = \sqrt{0.5}$ and $\alpha = 1$ so we can see that the values of $\alpha^2$ is bound in the range $[0.5, 1]$ for the great majority of the time. Note that, by definition, $\alpha$ cannot actually be greater than one so the time spent above 1 represents a statistical fluctuation arising from a finite ensemble size and similar remarks probably apply to the time spent below $\sqrt{0.5}$.}
\label{fig-powerrelate}
\end{figure}

Here we show numerical results of the correlation coefficient between the state deviations $\delta\rho_{\rm dV}$ and $\delta\rho_{\rm dW}$,
\begin{align}\label{eq-appcorr}
\alpha \equiv&\, 
\frac{{\cal R}_{\cal F}\dW}{\sqrt{{\cal R}_{\cal P}\dW {\cal R}_{\cal P}\dV}} 
=  \, \frac{ {\rm E}[\Tr[\delta\rho_{\rm dW}\delta\rho_{\rm dV}]] }{  \sqrt{{\rm E}[\Tr[(\delta\rho_{\rm dW})^2]]{\rm E}[\Tr[(\delta\rho_{\rm dV})^2]]}},
\end{align}
for all 27 cases of different dV and dW. In Figure~\ref{fig-powerrelate}, we show that $\alpha$ is always large for any combination of dOdVdW, and often much closer to 1.

\section*{References}

\bibliographystyle{iopart-num}

\providecommand{\newblock}{}

\end{document}